\documentclass{aa}
\usepackage{graphicx}
\usepackage{natbib}
\usepackage{amsmath}
\begin{document}
\authorrunning{Jurcsik et al.}
\title{The triple-mode pulsating variable V823 Cas}
\subtitle{
\thanks{Tables $1 - 5$ are only available in electronic form at the
CDS via anonymous ftp to cdsarc.u-strasbg.fr (130.79.128.5)
or via http://cdsweb.u-strasbg.fr/cgi-bin/qcat?J/A+A/}
}

\author{J.~Jurcsik\inst{1}, 
B.~Szeidl\inst{1},  M.~V\'aradi\inst{1}\inst{2}, A. Henden,\inst{3}
Zs.~Hurta\inst{4}, B.~Lakatos\inst{4}, K.~Posztob\'anyi\inst{4}, 
P.~Klagyivik\inst{4}, \and \'A. S\'odor\inst{4}
}

   \institute{Konkoly Observatory of the Hungarian Academy of Sciences. P.O.~Box~67,
H-1525 Budapest, Hungary\\
              \email{jurcsik@konkoly.hu}
   \and Department of Experimental Physics and Astronomical Observatory, University of Szeged,
H-6720 Szeged, D\'om t\'er~9, Hungary
\and Universities Space Research Association/U.S. Naval Observatory, P. O. Box 1149, Flagstaff, Arizona, U.S.A. 
                                                     86002-1149, U.S.A., e-mail: aah@nofs.navy.mil
   \and E\"otv\"os Lor\'and University, Department of Astronomy, P.O.~Box~32, H-1518
Budapest, Hungary
}

   \date{Received ; accepted }

\abstract{
Based on extended multicolour CCD photometry of the triple-mode radial 
pulsator V823~Cas we studied the properties of the coupling frequencies 
invoked by nonlinear processes. Our results support that a resonance connection 
as suggested by Antonello \& Aikawa (1998) affects the mode coupling behaviour. 
The P1/P0 period ratio of V823~Cas has an "out of range" value if compared
with the period ratios of the known double mode pulsators, while the P2/P1
period ratio is normal.
The periods and period ratios cannot be consistently interpret without
conflict 
with pulsation and/or evolution models. We attempt to interpret this
failure by the
suggestion that at present, the periods of V823~Cas are in a transient, resonance 
affected state, thus do not reflect the true parameters of the object. 
The anomalous period change behaviour of the fundamental and second overtone modes
supports this idea. We have also raised the possibility that a $f_0 +f_2 = 2f_1$ 
resonance may act in triple mode pulsators. 
 }  

   \maketitle


\section{Introduction}

Stellar pulsation often manifests multiperiodic behaviour. The 
simultaneous excitation of different modes is a common property
of e.g. $\delta$~Sct variables. In these stars both radial and 
nonradial pulsation modes can be excited, and especially in case 
of a dense frequency spectrum with many nonradial modes, the 
identification of the modes still has not been settled. 

Multiperiodicity of radial mode pulsators is also observed in
different types of variables. Radial mode pulsating variables are 
important targets of astero-seizmology as mode identification 
is unambiguous in many cases. Consequently, the pulsation properties 
of these stars reveal the physical parameters quite precisely by 
using pulsation models and/or empirical relations developed for the 
different types of variables along the instability strip. Two radial 
modes can be excited simultaneously in Cepheids, HADS (high
amplitude $\delta$~Sct stars) SX~Phe and RR~Lyrae variables as well. 
These, so called double mode pulsators, oscillate either in the 
fundamental and first overtone modes or in the first and second overtones.

In three stars; \object{AC~And}, \object{V823~Cas} and 
\object{V829~Aql}, the three lowest radial modes have
been detected 
to be excited simultaneously. Very recently two triple-mode short 
period Cepheids were also discovered in the LMC \citep{mp}, these 
stars oscillate, however, in the first three overtone modes. As a 
result of the mass photometry of the LMC and SMC (OGLE, MACHO, EROS 
projects, \citet{ud,sos,macho,beau}) dozens of short period 
($P<1$~d) double mode Cepheids were discovered, both fundamental 
and first overtone ($FU/FO$) and first and second overtone ($FO/SO$) 
pulsators. These results have also shown that in the LMC and SMC  most 
of the $P<1$~d Cepheids have double mode properties, indicating that in
their physical parameter regimes multi-mode pulsation is favoured. 
Because of the large distance and the consequent faintness of the 
LMC/SMC multi-mode pulsators, their investigations are strongly 
restricted. The three galactic triple-mode radial pulsators are 
also sparsely examined, in spite of that our knowledge on stellar 
pulsation would highly benefit from their thorough observational 
and/or theoretical studies. 

According to their fundamental periods \object{AC~And} and 
\object{V823~Cas} ($P_0\sim0.7$~d) resemble to metal poor RR~Lyrae 
stars, while \object{V829~Aql} with $P_0=0.29$~d is
classified as a 
post main sequence $\delta$~Sct star by \citet{handler}. 
The metal content of \object{AC~And} is, however, extremely high 
for a typical RR~Lyrae star \citep{preston}.  \citet{fsz} and 
\citet{kovacs} based on pulsation and evolution models have concluded 
that it is most probably a higher mass evolved star similar to 
$\delta$~Sct variables. \citet{fer2} suggested that \object{AC~And} 
may be the missing link between $\delta$~Sct stars and classical 
Cepheids. 

\object{V823~Cas} was discovered to be a triple mode radial pulsator 
by S.~Antipin \citep{antipin} using 
the Moscow photographic plate collection taken with the 40cm 
astrograph in Crimea between 1948 and 1995. Data showed without doubt light 
variation with three independent periodicities matching most probably 
to the three lowest radial mode oscillations. Until today, no further 
information on this important object has emerged, therefore, we 
decided to carry out an extensive multicolour photometric 
investigation of \object{V823~Cas}.

\section{Observations, data reduction}

The observations were obtained with the automated 60cm telescope of 
the Konkoly Observatory (Sv\'abhegy, Budapest) equipped with a 
Wright~750x1100~CCD (parameters and calibration are given in 
\citet{bakos}) using $BV(RI)_c$ filters. The field of view was 17'x24'.
Reduction processes were performed using standard 
{\sc IRAF}\footnotemark[1]\footnotetext[1]{{\sc IRAF} is distributed 
    by the National Optical Astronomy Observatories, which are operated 
    by the Association of Universities for Research in Astronomy, Inc., 
    under cooperative agreement with the National Science Foundation.} 
packages. Data were corrected for atmospheric extinction.

About 3\,600 frames were obtained in each passband on 38 nights
between 25~September and 14~December in 2003 
(JD~$2\,452\,908-2\,452\,988$). Transformation to the standard system 
was done using the $B,V,R_c,I_c$ magnitudes of surrounding stars 
observed by A. Henden with the USNO Flagstaff Station 1.0~m telescope 
equipped with a SITe/Tektronix~1024x1024~CCD. A complete table of 
the positions and $BV(RI)_c$ magnitudes of stars in the field centred 
on \object{V823~Cas}, is given in Table~\ref{henden} (available 
electronically from the CDS). The following formulae were derived to 
transform the instrumental data to the standard system:

$V=0.989v+0.095(b-v)$

$B=0.987b-0.032(b-v)$

$R_c=0.986r+0.112(v-r)$

$I_c=0.989i-0.012(v-i)$

Aperture photometry was applied to measure the relative magnitudes of
\object{V823~Cas} to \object{GSC~04018-01777} ($V=11\fm235, 
B-V=1\fm173, V-R_c=0\fm617, V-I_c=1\fm202$). \object{HD~134}, 
\object{GSC~04018-01891} and \object{GSC~04018-01661} were used as 
check stars, neither of them showed any significant variability relative 
to \object{GSC~04018-01777}. 
	
Photometric data are available electronically at the CDS. In 
Table~\ref{elB} $-$~\ref{elI} (http://cdsweb.u-strasbg.fr/cgi-bin/qcat?J/A+A/)
Column~1 lists the HJD of the observations, and Column~2 gives the
differential magnitudes of \object{V823~Cas} with respect to 
\object{GSC~04018-01777} for the $B,V, R_c, I_c$ colours.
$B-V$, $V-R_c$, and $V-I_c$ colour curves were also derived utilizing 
the $V$ observations and fitted values of the $B$, $R_c$, $I_c$ curves 
according the Fourier solutions given in Table~\ref{fouramp} and 
Table~\ref{fourphi} for the moments of the $V$ measurements. 
The colour indices are given in the 3.,4., and 5. Columns of Table~\ref{elV}.

\begin{table}
\caption{electronic table: BVRI magnitudes of calibrating stars in the field of V823 Cas}
\label{henden}
\end{table}
\begin{table}
\caption{electronic table: V823 Cas $B$ data}\label{elB}
\end{table}
\begin{table}
\caption{electronic table: V823 Cas $V, B-V, V-R_c, V-I_c$ data}\label{elV}
\end{table}
\begin{table}
\caption{electronic table: V823 Cas $R$ data}\label{elR}
\end{table}
\begin{table}
\caption{electronic table: V823 Cas $I$ data}\label{elI}
\end{table}

\section{Photometric results}

\subsection{Frequency components, amplitudes, and phases}
\begin{figure}[t]
   \centering
   \includegraphics[width=8.8cm]{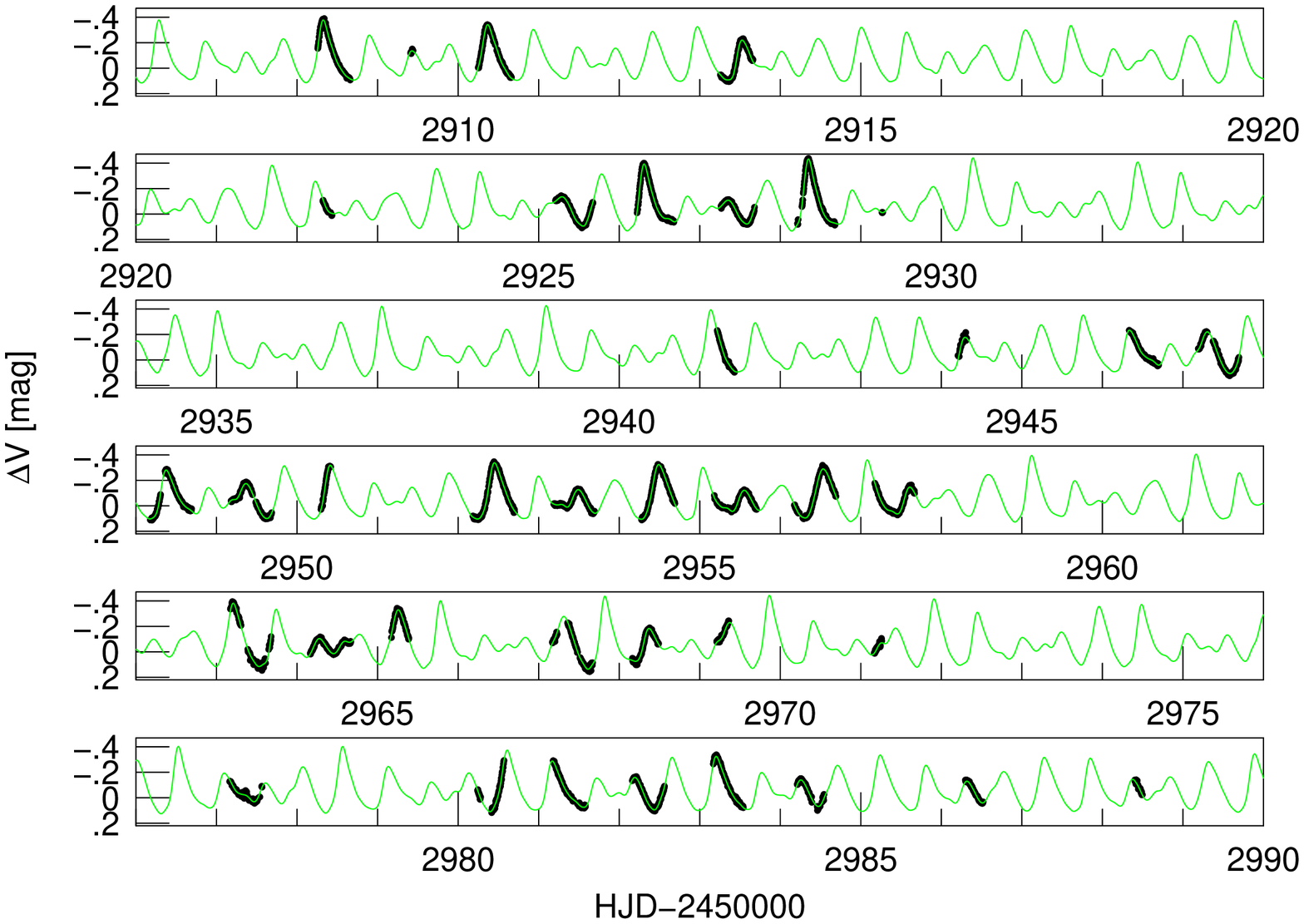}
   \caption{$V$ light curve of \object{V823~Cas} fitted with the Fourier 
solution given in Table~\ref{fouramp} and Table~\ref{fourphi}. 
}
              \label{fit}
\end{figure}
\begin{figure}[bbbhhh!!!!!!!]
   \centering
   \includegraphics[width=9.cm]{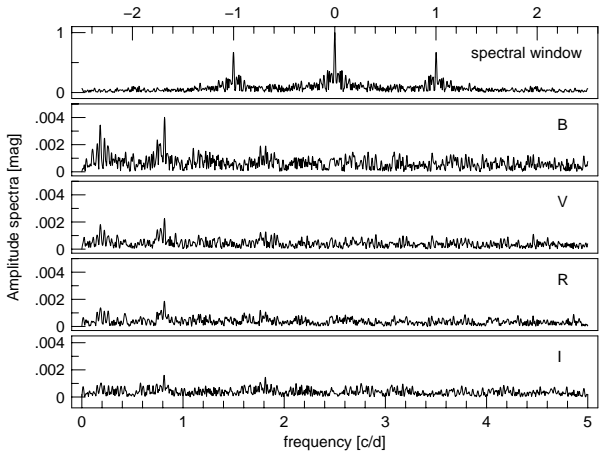}
   \caption{Residual spectra of the $B, V, R_c, I_c$ data after 
prewhitening with the frequencies of the three radial modes, their 
harmonics and the linear combinations of these frequencies appearing in 
the Fourier spectra. In each colour the highest signal appears at 
0.817~c/d indicating the detection of a real frequency component.
}
              \label{res}
\end{figure}

\begin{table*}[]
   \caption{Amplitudes of the detected frequencies in $B,V,R_c,I_c,B-V,V-R_c,V-I_c$.}
         \label{fouramp}
         \begin{tabular}{rrrrrrrrr@{\hspace{8pt}}r@{\hspace{8pt}}r}
            \hline\hline
            \noalign{\smallskip}

\multicolumn{3}{c}{Identification}&frequency&$A_B$&$A_V$&$A_{R_c}$&$A_{I_c}$&$A_{B-V}$&$A_{V-R_c}$&$A_{V-I_c}$ \\
            \noalign{\smallskip}
            \hline
            \noalign{\smallskip}
&&&cycle/day&\multicolumn{7}{c}{ mmag}\\
	    \noalign{\smallskip}
            \hline
            \noalign{\smallskip}
$-$&     $f_1$&$   -$&$  1.950515$&$  176.45$&$  130.23$&$  102.37$&$  79.59$&$  46.42$&$  28.01$&$  51.22$ \\ 
$-$&     $2f_1$&$  -$&$  3.901030$&$  37.15$&$  28.19$&$  21.97$&$  17.17$&$  8.77$&$  6.31$&$  11.09$  \\
$-$&     $3f_1$&$  -$&$  5.851545$&$  13.13$&$  10.25$&$  7.67$&$  6.96$&$  3.00$&$  2.54$&$  3.25$  \\
$-$&     $4f_1$&$  -$&$  7.802060$&$  3.62$&$  3.08$&$  2.52$&$  2.27$&$  0.50$&$  0.60$&$  1.11$  \\
$f_0$&   $-$&$     -$&$  1.494765$&$  115.95$&$  86.00$&$  67.10$&$  52.28$&$  30.17$&$  19.26$&$  34.60$ \\ 
$2f_0$&  $-$&$     -$&$  2.989530$&$  16.64$&$  12.28$&$  9.78$&$  8.09$&$  4.41$&$  2.49$&$  4.20$  \\
$3f_0$&  $-$&$     -$&$  4.484295$&$  2.14$&$  1.30$&$  0.88$&$  0.99$&$  1.31$&$  1.29$&$  1.31$  \\
$-$&     $-$&$   f_2$&$  2.433590$&$  29.16$&$  22.04$&$  17.67$&$  13.57$&$  7.21$&$  4.37$&$  8.47$ \\ 
$f_0$&   $f_1$&$   -$&$  3.445280$&$  44.77$&$  34.02$&$  26.97$&$  20.54$&$  11.23$&$  6.87$&$  13.22$ \\ 
$-f_0$&  $f_1$&$   -$&$  0.455750$&$  25.10$&$  19.45$&$  15.45$&$  12.15$&$  5.67$&$  4.02$&$  7.28$  \\
$2f_0$&  $f_1$&$   -$&$  4.940045$&$  11.76$&$  8.79$&$  7.30$&$  5.99$&$  3.17$&$  1.47$&$  2.74$  \\
$2f_0$& $-f_1$&$   -$&$  1.039015$&$  5.26$&$  3.72$&$  3.74$&$  2.68$&$  1.70$&$  0.44$&$  1.01$  \\
$3f_0$&  $f_1$&$   -$&$  6.434810$&$  3.44$&$  2.12$&$  1.56$&$  1.56$&$  1.28$&$  0.62$&$  0.77$  \\
$-$&     $f_1$&$ f_2$&$  4.384105$&$  10.11$&$  8.07$&$  6.31$&$  4.43$&$  2.29$&$  1.79$&$  3.73$  \\
$-$&     $-f_1$&$ f_2$&$  0.483075$&$  6.94$&$  5.43$&$  3.88$&$  3.35$&$  1.43$&$  1.64$&$  2.13$  \\
$f_0$&   $f_1$&$  f_2$&$  5.878870$&$  4.87$&$  3.18$&$  1.93$&$  2.39$&$  1.69$&$  1.40$&$  0.96$  \\
$-f_0$&  $f_1$&$  f_2$&$  2.889340$&$  4.16$&$  3.21$&$  2.73$&$  2.06$&$  0.93$&$  0.63$&$  1.21$  \\
$ f_0$&  $-f_1$&$ f_2$&$  1.977840$&$  4.16$&$  3.24$&$  2.52$&$  2.15$&$  1.19$&$  0.94$&$  1.55$  \\
$2f_0$&  $f_1$&$  f_2$&$  7.373635$&$  1.56$&$  1.52$&$  1.33$&$  0.76$&$  0.25$&$  0.27$&$  0.80$  \\
$f_0$&   $2f_1$&$  -$&$  5.395795$&$  10.70$&$  7.91$&$  5.97$&$  4.51$&$  2.78$&$  2.10$&$  3.41$  \\
$-f_0$&  $2f_1$&$  -$&$  2.406265$&$  13.50$&$  10.48$&$  8.67$&$  6.56$&$  2.97$&$  1.88$&$  4.00$  \\
$2f_0$&  $2f_1$&$  -$&$  6.890560$&$  3.82$&$  3.05$&$  2.27$&$  1.89$&$  0.68$&$  1.03$&$  1.31$  \\
$-2f_0$& $2f_1$&$  -$&$  0.911500$&$  2.07$&$  1.92$&$  1.25$&$  0.89$&$  0.17$&$  0.70$&$  1.24$  \\
$-$&     $2f_1$&$ f_2$&$  6.334620$&$  5.89$&$  4.18$&$  3.22$&$  2.61$&$  1.71$&$  0.97$&$  1.63$  \\
$f_0$&   $2f_1$&$ f_2$&$  7.829385$&$  2.52$&$  2.09$&$  1.56$&$  1.46$&$  0.42$&$  0.55$&$  0.95$  \\
$-f_0$&  $2f_1$&$ f_2$&$  4.839855$&$  2.11$&$  1.14$&$  0.73$&$  0.22$&$  1.31$&$  0.33$&$  0.76$  \\
$f_0$&   $3f_1$&$  -$&$  7.346310$&$  6.33$&$  4.80$&$  3.56$&$  3.20$&$  1.65$&$  1.19$&$  1.64$  \\
$-f_0$&  $3f_1$&$  -$&$  4.356780$&$  3.50$&$  3.15$&$  2.63$&$  1.55$&$  0.39$&$  0.65$&$  1.76$  \\
$2f_0$&  $3f_1$&$  -$&$  8.841075$&$  1.72$&$  1.58$&$  0.63$&$  0.53$&$  0.52$&$  0.95$&$  1.15$  \\
$-$&     $3f_1$&$ f_2$&$  8.285135$&$  3.05$&$  2.19$&$  1.27$&$  1.62$&$  1.00$&$  0.91$&$  0.59$  \\
$-$&     $3f_1$&$ -f_2$&$ 3.417955$&$  2.16$&$  1.57$&$  1.56$&$  1.29$&$  0.50$&$  0.19$&$  0.31$  \\
$f_0$&   $4f_1$&$  -$&$  9.296825$&$  2.89$&$  1.82$&$  1.65$&$  1.25$&$  1.19$&$  0.29$&$  0.58$  \\
$f_0$&   $-$&$   f_2$&$  3.928355$&$  6.01$&$  4.22$&$  3.17$&$  3.36$&$  1.88$&$  1.11$&$  0.80$  \\
$-f_0$&  $-$&$   f_2$&$  0.938825$&$  2.45$&$  2.22$&$  1.12$&$  1.16$&$  0.71$&$  1.12$&$  1.47$  \\
$f_3$&     $-$&$   -$&$  0.817000$&$  4.02$&$  2.27$&$  1.87$&$  1.43$&$  1.73$&$  0.49$&$  0.89$  \\
            \noalign{\smallskip}
            \hline
         \end{tabular}
\end{table*}

\begin{table*}[ttt!!!]
   \caption{Phases of the detected frequencies in $B,V,R_c,I_c,B-V,V-R_c,V-I_c$ using cosine
term Fourier sums and $T_0=2\,452\,908$ initial epoch value.}
         \label{fourphi}
         \begin{tabular}{rrrrrrrrr@{\hspace{8pt}}r@{\hspace{8pt}}r}
            \hline\hline
            \noalign{\smallskip}
\multicolumn{3}{c}{Identification}&frequency&$\Phi_B$&$\Phi_V$&$\Phi_{R_c}$&$\Phi_{I_c}$&$\Phi_{B-V}$&$\Phi_{V-R_c}$&$\Phi_{V-I_c}$
\\
            \noalign{\smallskip}
            \hline
            \noalign{\smallskip}
&&&cycle/day&\multicolumn{7}{c}{ deg}\\
	    \noalign{\smallskip}
            \hline
            \noalign{\smallskip}
$-$     &  $f_1$  & $-$    &$  1.950515$& $-$85.24 & $-$86.21 & $-$87.87 & $-$90.72 & $-$82.60 & $-$80.09 & $-$79.14  \\
$-$     & $2f_1$  & $-$    &$  3.901030$& 17.17    & 16.42    & 15.56    & 14.11    & 20.43    & 18.79    & 19.82     \\
$-$     & $3f_1$  & $-$    &$  5.851545$& 167.96   & 165.83   & 165.24   & 167.21   & 179.39   & 166.14   & 161.51    \\
$-$     & $4f_1$  & $-$    &$  7.802060$& $-$40.83 & $-$40.81 & $-$36.70 & $-$24.62 & $-$46.76 & $-$57.82 & $-$74.12  \\
$f_0$   & $-$     & $-$    &$  1.494765$& $-$9.53  & $-$11.77 & $-$14.50 & $-$18.37 & $-$3.11  & $-$2.26  & $-$1.81   \\
$2f_0$  & $-$     & $-$    &$  2.989530$& 220.48   & 218.75   & 217.74   & 215.09   & 225.52   & 222.46   & 225.57    \\
$3f_0$  & $-$     & $-$    &$  4.484295$& 69.47    & 66.47    & 57.45    & 58.57    & 247.29   & 66.61    & 66.60      \\
$-$     & $-$     & $f_2$  &$  2.433590$& $-$32.64 & $-$33.35 & $-$34.37 & $-$34.62 & $-$30.66 & $-$29.49 & $-$31.23  \\
$f_0$   & $f_1$   & $-$    &$  3.445280$& 156.53   & 156.86   & 157.49   & 157.50   & 156.46   & 154.16   & 155.79\\
$-f_0$  & $f_1$   & $-$    &$  0.455750$& 124.69   & 124.51   & 123.20   & 125.22   & 124.11   & 130.23   & 123.71    \\
$2f_0$  & $f_1$   & $-$    &$  4.940045$& 16.59    & 18.51    & 20.08    & 21.15    & 7.98     & 14.15    & 14.22\\
$2f_0$ & $-f_1$   & $-$    &$  1.039015$& 124.89   & 131.37   & 138.78   & 130.44   & 108.37   & 43.21    & 135.37\\
$3f_0$  & $f_1$   & $-$    &$  6.434810$& 259.64   & 248.62   & 244.81   & 234.59   & 275.08   & 259.32   & 277.82 \\
$-$     & $f_1$   & $f_2$  &$  4.384105$& 78.22    & 81.44    & 82.74    & 80.75    & 55.23    & 82.75    & 85.81     \\
$-$     & $-f_1$   & $f_2$ &$  0.483075$& 234.69   & 235.40   & 230.99   & 238.92   & 229.87   & 246.51   & 230.05\\
$f_0$   & $f_1$   & $f_2$  &$  5.878870$& $-$78.41 & $-$78.36 & $-$62.65 & $-$65.54 & $-$79.30 & $-$98.87 & $-$108.18    \\
$-f_0$  & $f_1$   & $f_2$  &$  2.889340$& 255.21   & 262.08   & 253.65   & 252.93   & 235.83   & 295.82   & 273.64  \\
$f_0$  & $-f_1$   & $f_2$  &$  1.977840$& 54.17    & 67.07    & 57.34    & 46.21    & 10.08    & 96.75    & 98.38\\
$2f_0$  & $f_1$   & $f_2$  &$  7.373635$& 119.60   & 131.29   & 137.83   & 145.01   & 44.71    & 91.55    & 116.02    \\
$f_0$   & $2f_1$  & $-$    &$  5.395795$& $-$80.14 & $-$77.50 & $-$70.74 & $-$72.81 & $-$88.61 & $-$96.49 & $-$83.44  \\ 
$-f_0$  & $2f_1$  & $-$    &$  2.406265$& 242.97   & 240.53   & 238.93   & 237.65   & 249.56   & 249.04   & 245.89    \\ 
$2f_0$  & $2f_1$  & $-$    &$  6.890560$& 158.83   & 158.21   & 171.31   & 170.19   & 169.90   & 126.06   & 138.32    \\
$-2f_0$ & $2f_1$  & $-$    &$  0.911500$& $-$3.69  & 2.88     & 3.42     & $-$27.58 & $-$77.19 & 0.01     & 22.05     \\ 
$-$     & $2f_1$  & $f_2$  &$  6.334620$& 224.63   & 224.04   & 228.02   & 217.36   & 225.35   & 209.85   & 234.81    \\
$f_0$   & $2f_1$  & $f_2$  &$  7.829385$& 95.76    & 91.42    & 94.31    & 114.91   & 112.82   & 84.46    & 55.10      \\
$-f_0$  & $2f_1$  & $f_2$  &$  4.839855$& 72.13    & 78.17    & 67.14    & 108.60   & 71.59    & 104.80   & 66.94     \\
$f_0$   & $3f_1$  & $-$    &$  7.346310$& 43.06    & 41.77    & 41.87    &  49.15   & 43.61    & 42.78    & 28.25     \\
$-f_0$  & $3f_1$  & $-$    &$  4.356780$& 30.46    & 39.13    & 33.64    & 38.47    & $-$44.35 & 54.05    & 35.88     \\
$2f_0$  & $3f_1$  & $-$    &$  8.841075$& $-$81.99 & $-$67.51 & $-$65.18 & $-$37.29 & $-$147.19& $-$68.90 & $-$79.12  \\
$-$     & $3f_1$  & $f_2$  &$  8.285135$& 29.87    & 20.70    & 18.68    & 21.75    & 54.73    & 21.53    & 13.96     \\
$-$     & $3f_1$  & $-f_2$ &$  3.417955$& 34.45    & 48.91    &  52.36   & 43.12    & $-$1.11  & $-$12.86 & 58.26     \\
$f_0$   & $4f_1$  & $-$    &$  9.296825$& 199.65   & 185.55   & 178.67   & 191.70   & 218.01   & 236.97   & 175.97    \\
$f_0$   & $-$     & $f_2$  &$  3.928355$& 134.81   & 135.73   & 141.33   & 137.49   & 140.06   & 114.45   & 123.12    \\
$-f_0$  & $-$     & $f_2$  &$  0.938825$& 171.24   & 155.66   & 169.61   & 197.53   & 222.10   & 141.50   & 124.41    \\
$f_3$     & $-$   & $-$    &$  0.817000$& 76.96    & 82.20    & 75.96    & 76.14    & 68.92    & 108.98   & 92.31
\\
            \noalign{\smallskip}
            \hline
		         \end{tabular}
\end{table*}

\begin{table}[hhhh]
   \caption{Generalized phase differences and amplitude ratios of the frequencies of the $B$ 
light curve solution:\newline
$G_{ijk}=\phi_{ijk}-(i\phi_{100}+j\phi_{010}+k\phi_{001})$, \newline
$R_{ijk}=(|i|+|j|+|k|) A_{ijk} / (|i|A_{100}+|j|A_{010}+|k|A_{001})$.
}
         \label{gij}
\begin{tabular}{@{\hspace{1pt}}c@{\hspace{1pt}}r@{\hspace{2pt}}r@{\hspace{2pt}}rcc@{\hspace{5pt}}c@{\hspace{1pt}}c@{\hspace{1pt}}c}
            \hline\hline
            \noalign{\smallskip}
order&$i$&$j$&$k$
& frequency & $A(B)$&$R_{ijk}$ &Error($\Phi_{ijk}$)& $G_{ijk}$ \\
            \noalign{\smallskip}
            \hline
            \noalign{\smallskip}
1&0&      1&       0&       1.950515&  0.1765& 1.000&  0.002& 0.000\\
&1&       0&       0&       1.494765&  0.1160& 1.000&  0.003& 0.000\\
&0&       0&       1&       2.433590&  0.0292& 1.000&  0.011& 0.000\\
            \noalign{\smallskip}
            \hline
            \noalign{\smallskip}
2&1&	1&	0&	3.445280&	0.0448& 0.307& 0.008&	4.384\\
&0&	2&	0&	3.901030&	0.0372& 0.211& 0.008&	3.273\\
&$-1$&	1&	0&	0.455750&	0.0251& 0.172& 0.013&	3.496\\
&2&	0&	0&	2.989530&	0.0166& 0.143& 0.020&	4.179\\
&0&	1&	1&	4.384105&	0.0101& 0.099& 0.035&	3.421\\
&0&	$-1$&	1&	0.483075&	0.0069& 0.068& 0.059&	3.176\\
&1&	0&	1&	3.928355&	0.0060& 0.083& 0.055&	3.087\\
&$-1$&	0&	1&	0.938825&	0.0025& 0.172& 0.149&	3.390\\
	    \noalign{\smallskip}
            \hline
            \noalign{\smallskip}
3&$-1$&	2&	0&	2.406265&	0.0135& 0.087& 0.025&	0.763\\
&0&	3&	0&	5.851545&	0.0131& 0.075& 0.026&	1.108\\
&2&	1&	0&	4.940045&	0.0118& 0.086& 0.028&	2.109\\
&1&	2&	0&	5.395795&	0.0107& 0.069& 0.031&	1.742\\
&0&	2&	1&	6.334620&	0.0059& 0.046& 0.059&	1.179\\
&2&	$-1$&	0&	1.039015&	0.0053& 0.039& 0.063&	1.024\\
&1&	1&	1&	5.878870&	0.0049& 0.046& 0.069&	0.855\\
&$-1$&	1&	1&	2.889340&	0.0042& 0.039& 0.090&	0.059\\
&1&	$-1$&	1&	1.977840&	0.0042& 0.039& 0.080&	0.194\\
&3&	0&	0&	4.484295&	0.0021& 0.018& 0.180&	1.711\\
	    \noalign{\smallskip}
            \hline
            \noalign{\smallskip}
4&1&	3&	0&	7.346310&	0.0063& 0.039& 0.054&	5.378\\
&2&	2&	0&	6.890560&	0.0038& 0.026& 0.087&	6.077\\
&0&	4&	0&	7.802060&	0.0036& 0.021& 0.093&	5.236\\
&$-1$&	3&	0&	4.356780&	0.0035& 0.022& 0.095&	4.826\\
&3&	1&	0&	6.434810&	0.0034& 0.026& 0.094&  6.510\\
&0&	3&	1&	8.285135&	0.0031& 0.022& 0.103&	5.551\\
&1&	2&	1&	7.829385&	0.0025& 0.020& 0.124&	5.380\\
&0&	3&	$-1$&	3.417955&	0.0022& 0.016& 0.160&	4.492\\
&$-1$&	2&	1&	4.839855&	0.0021& 0.017& 0.162&	4.635\\
&$-2$&	2&	0&	0.911500&	0.0021& 0.014& 0.150&	2.577\\
&2&	1&	1&	7.373635&	0.0016& 0.014& 0.211&	4.475\\
	    \noalign{\smallskip}
            \hline
            \noalign{\smallskip}
5&1&	4&	0&	9.296825&	0.0029& 0.018& 0.111&	3.314\\
&2&	3&	0&	8.841075&	0.0017& 0.011& 0.193&	3.363\\
            \noalign{\smallskip}
            \hline
		         \end{tabular}
\end{table}

The light curve solution given in Table~\ref{fouramp} and 
Table~\ref{fourphi} was determined from the Fourier analysis of the 
photometric data using the utilities of the  program package MUFRAN 
\citep{mufran}. Nonlinear regression facilities of Mathematica (Wolfram 
Research, Inc) were also applied. The listed frequencies fit the 
$B,V,R_c,I_c$ data with 0.0126, 0.0079, 0.0081, 0.0079 mag r.m.s. scatter, 
respectively, which is about the observational accuracy limit. The 
$V$ light curve and the fit are shown in Fig.~\ref{fit}.

Besides the frequencies of the three dominant radial modes and their 
harmonics, 26 of the possible linear combination coupling frequencies 
could be identified in the successively prewhitened spectra of the $B$ data. 

Frequency component identification was performed on the $B$ dataset as
the amplitudes of the signals are the largest in $B$. In order to refine 
the $f_0, f_1, f_2$ frequency values least squares solutions were 
calculated within the vicinity of the suspected values of $f_0, f_1, f_2$, 
their harmonics and the detected linear combination term frequencies. 
Linear combination frequencies were always set according to the actual 
values of the main frequency components. Solution with the smallest 
residual scatter was accepted for the $f_0, f_1, f_2$ frequency values. 
The $2\sigma$ asymptotic error estimates of the corresponding periods 
are 0.000045, 0.000013 and 0.000060~days, respectively.

The formal errors ($1\sigma$) of the amplitudes are about $0.2-0.4$~mmag, the errors 
of the phases of signals with amplitude larger than 1 mmag are reliable 
within $0.1-10$\degr.

All but $3f_1-2f_0$ 2-term coupling frequencies involving 
$f_1, 2f_1, 3f_1,$ and $f_0$, $2f_0$ appear in the spectra, and six 
3-term linear combination frequencies are also detected. The amplitudes 
of the coupling components are always smaller than that of its compounding 
frequencies with one exception. The amplitude of the $f_1+3f_0$ 
component is higher than the amplitude of the $3f_0$ in each colour.
Actually the $3f_0$ harmonic component is only marginally present in 
the data, we involved this component to the solution only because a real 
frequency component was found at 6.43481~c/d, which we could interpret 
only as $f_1+3f_0$. However, considering the marginal amplitude of the 
$3f_0$ component, this interpretation might not be correct.

After the removal of the radial modes, their harmonics and the coupling 
terms from the data, the residual spectra in each colour show evidence 
of a component at 0.817~c/d as shown in Fig.~\ref{res}. This frequency, 
denoted as $f_3$ in Table~\ref{fouramp} and Table~\ref{fourphi}, cannot 
be resolved as the linear combination of the radial modes. As no 
indication of this frequency was found in the comparison - check stars 
data, we suppose it as a real frequency component of \object{V823~Cas}.
Taking into account the small amplitude and the relatively long, 1.22~d 
periodicity of this variation one may speculate about its origin as a 
high-order gravity mode, similarly as in the $\gamma$~Dor variables. In 
order to check the reality of this explanation, however, further 
observational and theoretical confirmations are needed.

No other frequency in the residual spectrum was detected, consequently 
the solution given in Table~\ref{fouramp} and Table~\ref{fourphi} is a 
full description of the data within the limits of the accuracy of the 
observations.

In most cases the linear combination components involving negative 
term have smaller amplitudes than the combination component of the 
same but positive terms, in agreement with model results, that 
nonlinear pulsation favours combination frequency terms where both 
$i$ and $j$ are positive \citep{ana}. There is one significant 
counter-example, $2f_1-f_0$ has larger amplitude than $2f_1+f_0$ in 
each colour.
The amplitudes of the 3-term combination frequencies of the same 
components have similar amplitudes independently of the signs.

\begin{figure}[t]
 \centering
   \includegraphics[width=6cm]{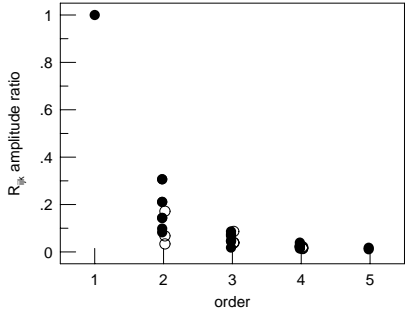}
   \caption{Amplitude ratios of the different order combination frequency terms.
Filled and open circles denote frequencies with positive $i,j,k$ values
and involving negative terms as well, respectively. The decrease in the
amplitude ratios is exponential with increasing order.
 }
              \label{rij}
\end{figure}

The amplitude ratios of the combination frequencies defined as
\newline
$R_{ijk}=(|i|+|j|+|k|) A_{ijk} / (|i|A_{100}+|j|A_{010}+|k|A_{001})$
\newline
decrease exponentially with increasing order ($|i|+|j|+|k|$) as shown in 
Fig~\ref{rij}. $A_{ijk}$ denotes the amplitude of the $if_0+jf_1+kf_2$ frequency
component, i.e., $A_{100}=A(f_0), A_{010}=A(f_1)$, and $A_{001}=A(f_2)$.

The amplitudes of all the detected frequencies decrease towards longer 
wavelengths. Most of the frequencies of the light curve solution 
appear in the colour indices as well, typically with the largest 
amplitude in the $V-I_c$ colour.

The phases of the three pulsation modes tend to be smaller towards the 
longer wavelengths. The phase differences between the $B$ and $I_c$ 
data for $f_0$, $f_1$ and $f_2$ are $9\degr$, $5\degr$, and $2\degr$, 
respectively, while the phases of the colour indices are typically 
larger than the phases of the light curves, e.g. the phases of $f_0$, 
$f_1$ and $f_2$ of the $B-V$ curve are larger by about $8.7\degr$, 
$3.6\degr$, and $2.7\degr$, than the phases of the $V$ light curve
solution. In contrast with these systematic trends of the phases of 
the three radial modes, the phases of the two largest amplitude linear 
combination terms, $f_1+f_0$ and $f_1-f_0$, are identical in each 
wavelength bands. The phases of the smaller amplitude linear 
combination terms are not accurate enough to detect any real wavelenght 
dependent trend in the phase behaviour.  

\citet{pp} have shown that in galactic double-mode Cepheids the 
different order generalized phase differences ($G_{ijk}$, for a 
definition see \citet{a}) of the linear combination term frequencies
tend to have similar values. In Table~\ref{gij} the different order 
generalized phase differences of our $B$ light curve solution and also 
the $R_{ijk}$ amplitude ratios are listed, in the order of decreasing 
amplitude. The $1\sigma$ errors of the phases are also given, in order 
to easily estimate the reliability of the $G_{ijk}$ phase difference values. 
For compatibility with other 
similar data, the phase differences are given in radians. 
The averages of the $G_{ijk}$ 
phase differences are 3.55, 1.07, 5.01 (5.26 omitting the most deviant 
2.577 value), and 3.34 for the 2., 3., 4. and 5. order terms, 
respectively. These values are systematically smaller by about 
$0.5-1.0$~rad as the corresponding 4.31, 2.17, 6.24, 3.85 averages of 
the $G_{ij}$ values of galactic double-mode Cepheids calculated from the
data listed by \citet{pp}. Taking into 
account the decreasing trend of the phase differences with decreasing 
periods as shown in Fig. 2 in \citet{pp} the short periods of 
\object{V823~Cas} may account for its smaller $G_{ijk}$ values.

\begin{figure}[]
   \centering
   \includegraphics[width=8.8cm]{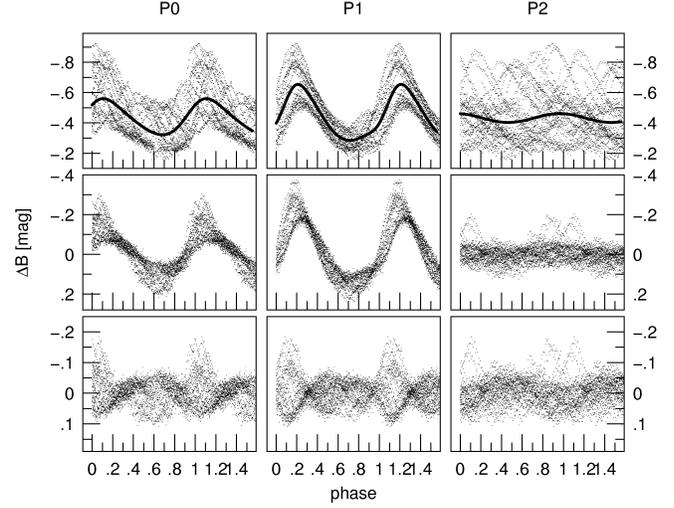}
   \caption{$B$ light curve of \object{V823~Cas} folded with the
periods of the three radial modes (top panels). The synthetic light
curves corresponding to the three periods are also shown calculated 
from the amplitudes and phases of the three radial modes and their 
harmonics according to the data given in Table~\ref{fouramp} and 
Table~\ref{fourphi}. Middle panels show the folded light curves after 
subtracting the other two radial modes and their harmonics. The shape
of the light curve of a given mode is strongly variable 
depending on the relative phases of the modes. In the bottom panels
residual data after removing the light curves of the three radial 
modes are shown. These panels clearly indicate that the amplitude of 
the light variation increases significantly when each of the modes are 
simultaneously around maximum brightness. This enhancement
may account for the numerous coupling frequencies appearing 
in the Fourier spectrum of the data in accordance with the model results 
of \citet{ana}.
}
              \label{phase}
\end{figure}

As indicated by \citet{ana} the only plausible explanation that can 
change the phases and can lead to coherent values of the different 
order phase differences is a resonance between the pulsation modes. In 
Fig~\ref{phase} the $B$ light curve is folded according to the periods 
of the three radial modes (top panels), after prewhitening with the 
frequencies and harmonics of the other two modes (middle panels), and 
the residual light curve after the removal of all the three radial 
modes' light curves (bottom panels). These plots still show some kind 
of regular behaviour according to the main periodicities. It is also 
evident that extreme large amplitudes occur when each of the modes 
are simultaneously around maximum phase indicating also that some 
kind of resonance interaction may be responsible for the occurrence 
of the coupling frequencies.

\begin{figure}[tthhh!!!]
   \centering
   \includegraphics[width=8.5cm]{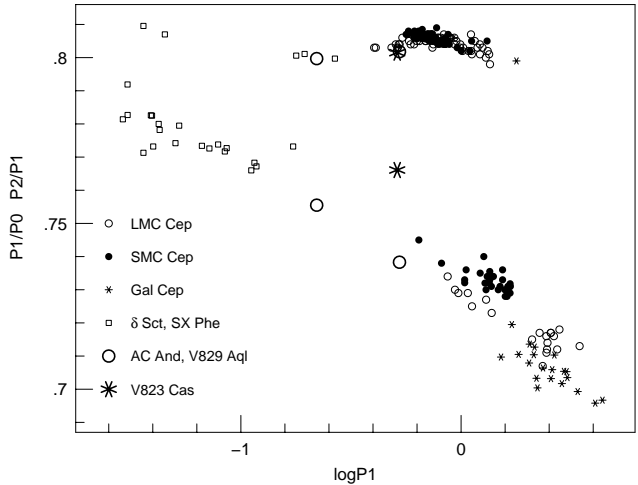}
   \caption{Period ratios as a function of $\log P_1$ of different types of 
multi-mode radial pulsators. Though $P_1$ is actually the same for AC~And
and V823~Cas, their $P_1/P_0$ period ratios differ about 0.03 which cannot
be explained with any major difference in their physical parameters 
consistently with pulsation and evolutionary models.
}
              \label{dm}
\end{figure}
\begin{figure}
   \centering
   \includegraphics[width=8.8cm]{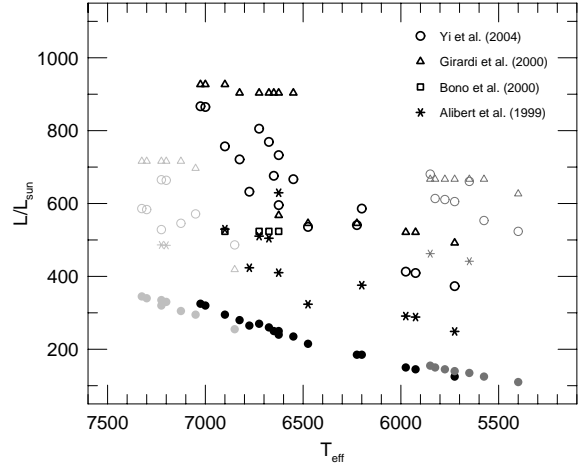}
   \caption{Comparison of pulsation solutions for the observed periods and period ratios
of V823 Cas with evolutionary model results. Filled circles denote pulsation solutions
according to the KB94 models for Z=0.004 (light gray), Z=0.01 (black) and Z=0.02 (dark
gray) compositions, respectively.
Other symbols indicates the luminosities of evolutionary tracks during first crossing of
the same
mass and metallicity models at the temperature of the pulsation solutions.
Second, third crossing luminosities are at even larger luminosities.
The corresponding evolutionary models have the same  mass within $ \pm 0.25$~M$_{\sun}$ 
and the same metallicity within $ Z=\pm0.002$ ranges as the pulsation model solutions
have.
Each of the evolutionary models indicates larger luminosity
than the pulsation solution. 
  }
              \label{model}
\end{figure}

\section{Comparison with model results}

Period ratios of different type radial multi-mode 
pulsators has been collected and the result is summarized in Fig~\ref{dm}. 
The $\sim 0.01$ separations of 
the $P_1/P_0$ sequences of galactic, LMC and SMC Cepheids reflect the 
different metal content of the stars ($Z=0.02, 0.008$ and 0.004, 
respectively). The $P_1/P_0$ ratio of 
\object{V823~Cas} is evidently discrepant from the global $\log P_1 - P_1/P_0$ 
relation defined by the different types of variables.
This is most probable because of an anomalous value of $P_0$, as the $P_2/P_1$ 
period ratio fits well the observed values of other multi-mode pulsators.    

In order to find an answer for the anomalous value of the fundamental mode period
first we have checked whether the observed periods and period ratios 
can match stellar pulsation and evolution model results.
The difference between pulsation periods  derived from 
linear models and the real nonlinear periods are supposed to be negligible.
Thus linear periods are widely used to determine physical parameters of multi-mode 
pulsators (e.g. \citet{mp,pc,ana}). \citet{kovacs} (hereafter KB94)
calculated an extended 
set of linear radial models to explain the periods of \object{AC~And}, 
the other triple-mode variable with very similar periods.  
We have compared the pulsation periods calculated from the KB94 formulae with those
given in the evolutionary and pulsation models of \citep{alibert}.
The period of a given model agrees within 1\% accuracy with the period calculated from
KB94 models (within the common parameter regimes) indicating consistency of the 
two model groups.  
 
The observed periods, and period ratios $P_0=0.669001$~d, $P_1/P_0=0.7663, P_2/P_1=0.8015,$ 
can be fitted with period ratio distance $\Delta < 0.003$ (for definition see KB94)
accuracy with certain high-mass-sequence~2 models of KB94.
Solutions can be found for the Z=0.004 and Z=0.01 models  
with $3.7-4.0$ M$_{\sun}$ and $3.85 - 4.25$ M$_{\sun}$ mass, respectively.
The metal-poor solutions ($Z=0.004$) are at too high,
$T>7\,000~\mathrm{K}$ temperature 
out of the instability strip, with negative growth rates 
of the detected modes, therefore we exclude these among the real possibilities.
There is also indication that solutions might also exist for solar composition, 
Z=0.02 models, but at 'out of range' $M\sim 4.5$ mass value, which   
was not covered with the KB94 models, thus these pulsation solutions are somewhat
uncertain.
Moreover, considering the Period - Temperature connections valid for Cepheids, e.g.,
Fig.~1 in \citet{ba} and Fig.~3 in  \citet{sandage}, the temperature of
Cepheids
with
periods shorter than 1~day must be higher than 6000 K. All the Z=0.02 model period
solutions fall below 6000 K, thus these are not valid solutions as well. 

The parameter regimes of the possible solutions are shown in Fig.~\ref{model}.
The physical parameters of these models seems to be, however, in 
conflict with evolutionary results.
At a given mass and chemical composition any canonical evolutionary model predicts the 
possible lowest luminosity within the instability strip during its
'first crossing', evolving off the main sequence. 
However, each of the appropriate evolutionary models
\citep{alibert,bono,girardi,yy}
cross the instability strip
at significantly larger  luminosities than the
solutions of the
periods
indicate even during the first crossing.
The minimal luminosity at a given metallicity and mass values according to evolutionary
models are also indicated in Fig.~\ref{model}.
The $100-600$~L$_{\sun}$ discrepancy between luminosities allowed by stellar evolution 
and luminosities derived from the pulsation solution of the periods indicates also 
that the periods of V823 Cas are anomalous. 

This problem arises from the unusually large value observed for the 
$P_1/P_0$ ratio (0.766). At $P_0\approx0.7$~d the normal value of $P_1/P_0$ 
would be less than 0.75 for any consistent model. 
The extended linear nonadiabatic pulsation models for galactic, LMC and SMC
compositions \citep{morgan} strengthen that  $P_1/P_0$ can never be larger than 0.75
in any real case.
There are two ways to increase the period ratio at a given period,
towards more metal poor models, and/or with models of smaller $L/M$ ratio
the $P_1/P_0$  ratio is larger \citep{pc,kovacs}.
For \object{V823~Cas}, however, 
both of these possibilities can be excluded, as already mentioned the metal poor
period solution fall out of the instability strip, and small $L/M$ ratio is not allowed 
by evolutionary models.

As a conclusion, we have found that the period ratios of \object{V823~Cas} 
cannot be explained consistently with stellar evolution and pulsation 
models. We suppose that this is not the consequence of any serious
defect/inadequacy of the models but more probably arises from a temporal,
transition behaviour of \object{V823~Cas}, which produces unusual period 
ratios (see the next section).

\section{Period changes}

\begin{table*}[]
   \caption{Periods and period change rates of \object{V823~Cas} and \object{AC~And}}
         \label{perc}
\begin{tabular}{@{\hspace{-1pt}}l@{\hspace{-2pt}}c
@{\hspace{10pt}}c@{\hspace{5pt}}c@{\hspace{0pt}}c
@{\hspace{1pt}}c@{\hspace{7pt}}c@{\hspace{0pt}}c
@{\hspace{1pt}}c@{\hspace{-1pt}}c@{\hspace{-5pt}}c@{\hspace{-0pt}}c}
            \hline\hline
            \noalign{\smallskip}
star&JD-2\,400\,000
& $P_0$ [d]&2$\sigma(P_0)$ &$Amp(P_0)$
& $P_1$ [d]&2$\sigma(P_1)$ &$Amp(P_1)$
& $P_2$ [d]&2$\sigma(P_2)$ &$Amp(P_2)$ &\\
            \noalign{\smallskip}
            \hline
            \noalign{\smallskip}
\multicolumn{2}{l}{V823 Cas}&&&&&&&&&&\\
\noalign{\smallskip}
            \hline
            \noalign{\smallskip}
&$52\,908-52\,988$&0.669001&.000045&0.12&0.512685&.000013&0.18&0.410916&.000060&0.03&CCD(B) (1)\\

&$42\,273-49\,033$&0.668887&.000005&0.13&0.512660&.000002&0.18&0.411027&.000005&0.05&pg (2)\\

&$37\,575-42\,016$&0.668870&.000009&0.12&0.512646&.000003&0.17&0.411030&.000007&0.05&pg (2)\\

&$29\,076-36\,540$&0.668840&.000009&0.13&0.512628&.000003&0.17&0.411040&.000007&0.07&pg (2)\\
            \noalign{\smallskip}
            \hline
            \noalign{\smallskip}
\multicolumn{2}{l}{AC And}&&&&&&&&&&\\
\noalign{\smallskip}
            \hline
            \noalign{\smallskip}
      &$47\,864-49\,039$&  0.711253&.000014&&0.525155&.000009&&0.421088&.000014&&Hip (3)\\
      &$36\,459-37\,949$&  0.711232&.000002&&0.525133&.000001&&0.421072&.000002&&pe(V) (4)\\
      &$30\,258-35\,009$&  0.711227&.000001&&0.525130&.000001&&0.421067&.000001&&pg (5)\\
      &$25\,540-27\,278$&  0.711217&.000011&&0.525108&.000008&&0.421063&.000016&&vis (6)\\
            \noalign{\smallskip}
            \hline
            \noalign{\smallskip}
\multicolumn{2}{l}{Period change rates} &${\dot{P_0}/{P_0}}$&&
&${\dot{P_1}/{P_1}}$&&
&${\dot{P_2}/{P_2}}$ &[$Myear^{-1}$]&&\\
            \noalign{\smallskip}
            \hline
            \noalign{\smallskip}
V823 Cas&&  increasing&&&  2.17 &&&decreasing&	 \\
AC And&&	 0.71 &&&  0.79 &&& 1.08&	 \\
	    \noalign{\smallskip}
            \hline
		         \end{tabular}
{\footnotesize
\underline {References of the data:}
(1) this paper; (2) \citet{antipin}; (3) \citet{hip}; (4) \citet{fsz}; (5) \citet{pg}
(6) \citet{vis1,vis2,vis3}
}
\end{table*}

The extended photographic data complemented with the recent CCD 
observations enable us to derive reliable period changes of the 
three observed modes. \citet{antipin} analysed  two sets of the 
photographic data (between $2\,432\,853-2\,439\,145$ and 
$2\,440\,071-2\,449\,633$, respectively) and detected period 
increase of the fundamental and first overtone modes. As photographic
data are available already from JD 2\,429\,076, we divided the 
photographic data into three subsets in order to get more details of 
the period change behaviour of \object{V823~Cas}. Table~\ref{perc} 
lists the periods of the radial modes, their $2\sigma$ error estimates
and the photographic or CCD($B$) amplitudes derived from nonlinear 
regression process of Mathematica.

For comparison, the period change behaviour of \object{AC~And} has 
been revised and listed in Table~\ref{perc}, too. Periods for the different 
datasets of \object{AC~And} has been determined similarly as described in 
Sect 3.1.

\begin{figure}[hhh]
   \centering
   \includegraphics[width=9.2cm]{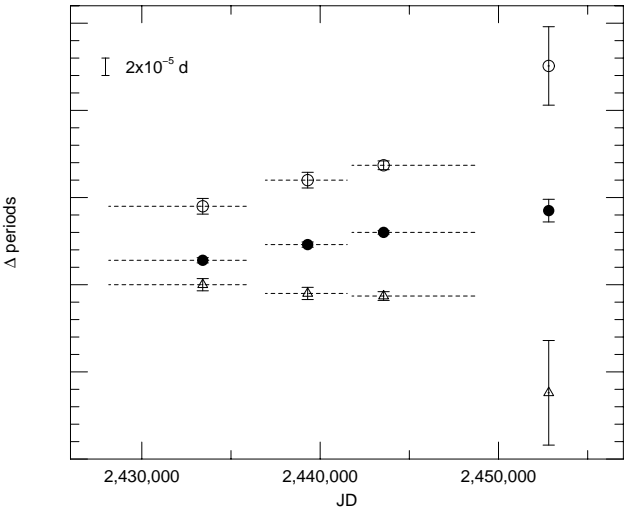}
   \caption{
Detected periods of \object{V823~Cas} at different epochs.
$P_0-0.66854$~d ($\circ$), 
$P_1-0.51240$~d ({\large{$\bullet$}}),
and $P_2-0.41088$~d ($\triangle$) are plotted for clarity.
}
              \label{per}
\end{figure}

The period change rate of the dominant, first overtone mode of 
\object{V823~Cas} and that of all the three modes of \object{AC~And} 
well correspond to the theoretically expected evolutionary value
during the first crossing of the instability strip. The evolutionary 
period change rates during the second and third crossing are by two orders 
of magnitude slower \citep{mp}.
The larger $\dot{P_1}/{P_1}$  indicates somewhat larger mass
of \object{V823~Cas} than AC~And.

In contrast, the periods of the fundamental mode and the second overtone 
of \object{V823~Cas} behave differently. According to the CCD 
observations the period change rates of these modes have been recently 
changed, $P_0$ has increased, while $P_2$ has decreased by about 0.0001~d 
during the last decades. 
The amplitude of the second overtone also shows significant 
changes, it has gradually decreased from 0.07~mag to 0.03~mag, in 
contrast with the amplitudes of the fundamental and first overtone modes, 
which differ only within the limits of the uncertainty for the 
different sets of the observations. The reduction of the $P_2$ amplitude 
indicates that this mode may be just diminishing, and this is 
accomplished by the extreme period change behaviour. However, this 
effect cannot be solely account for the rapid period increase of $P_0$. 
As we have no reason to assume that canonical evolution and 
pulsation models which fail to interpret the period ratios of 
\object{V823~Cas} were in fault, we suggest that, instead, the 
fundamental mode period is  somehow 'wrong'.

An interesting possibility arises from the opposite sign period change 
of the second overtone mode. Extrapolating backward the linear period
changes detected between JD~$2\,430\,000 - 2\,445\,000$, the period ratios should 
have been about $P_1/P_0=0.76679$ and $P_2/P_1=0.81089$ about 3\,300
years ago.
These period ratios fulfill the $f_0+ f_2=2f_1$ resonance criterion, as
$f_0=0.76679 f_1$ and
$f_2=1.23321 f_1$, indicating that the three radial modes might
have been in a resonance connection in a very recent past in astronomical
time-scale.

This resonance hypothesis would also help to understand the anomalous 
period ratios of \object{V823 Cas}. If we assume that in the 
future the period change rate of the first overtone remains the same 
as detected for the last 50 years, but the period increase
of the fundamental keeps on to be larger than that as indicated by recent
observations, i.e.,

$\dot{P_1}/P_1-\dot{P_0}/P_0= C < 0, \,\,\, C\sim const.$

\noindent
then, this would lead to change in the period ratio, as

${{d(P_1/P_0)}\over{dt}} = P_1/P_0\times(\dot{P_1}/P_1 -\dot{P_0}/P_0)=C\times
P_1/P_0.$

\noindent
Consequently, with time elapsing,

$(P_1/P_0)_t = (P_1/P_0)_{t=0}\times e^{Ct}$.

\noindent
The exponential change in the period ratio results drastic
changes during astronomically very short time base. 
The recent observations have shown that the period change 
rate of $P_0$  has increased to 3 times that of the period 
change rate of $P_1$. Assuming the possible 
largest uncertainties, $C$ is smaller than $-2\,Myear^{-1}$ at 
present, which, according to the above formula would lead to a 
decrease in the $P_1/P_0$ ratio from its unusual 0.766 to a normal 
0.745 value within 14\,000 years. 

\section{Absolute parameters derived from observed magnitudes, and P-L
relations}

The magnitude and intensity mean $V$ brightness and colours of 
\object{V823~Cas} using the standard magnitudes of the comparison 
star as given in Section~2 are the followings.
\noindent
Magnitude mean values: 

\noindent
$V=11.175, B-V=0.792, V-R_c=0.473, V-I_c=0.987$.

\noindent
Intensity mean values: 

\noindent
$V=11.169, B-V=0.787, V-R_c=0.471, V-I_c=0.982$.

The interstellar reddening in the direction of \object{V823~Cas} 
according to the \citet{schlegel} maps is quite large, $E(B-V)=0.887$~mag.
\object{V823~Cas} lies close to the galactic plane ($b=0.99\degr$) to the 
direction of the Perseus Arm. The distance of the Perseus Arm is about 
3~Kpc in the direction of \object{V823~Cas} ($l=118\degr$) as shown in 
Fig.~2 in \citet{q}. Consequently, it can be assumed that within 3~Kpc 
interstellar reddening is much smaller than 0.887~mag.
Dereddened colours, using  standard reddening law for
$BV(RI)_c$ colours as given in \citet{schlegel} fit the synthetic colours of
atmosphere models \citep{kurucz} according to the parameters listed in
Table~\ref{par}.

The observed periods of \object{V823~Cas} are too short for 
classical Cepheids and are too long for $\delta$ Scuti models. This
intermediate status makes comparison with model results uncertain, 
as in most of the cases conclusions can be drawn from extrapolation.
Because of this deficiency and the uncertainties 
of data conversion between the observable (magnitude, colour-indices) and the
theoretical plane's (luminosity, temperature) quantities,
the absolute parameters from the observed colours and periods 
cannot be determined accurately. 

A rough estimate can be, however, done by combining the possible solutions 
listed in Table~\ref{par} with predictions of empirical and theoretical  Period -
Luminosity relations.
The 0.669 d fundamental mode, and 0.513 first overtone periods correspond to $M_V=-0.9$
~mag absolute visual magnitude during first crossing
within $\pm0.3$~mag range for $Z=0.02-0.004$ matallicity models according to the
Period - Luminosity relations of \citet{ba} and \citet{alibert}. 
This value is in accordance with combined empirical $P-L$ relations of
Cepheids and $\delta$ Scuti stars \citep[e.g.][]{fer1}. 
Recent empirical studies e.g., \citet{udalski,
sandage} yield also consistent results within the given magnitude range.

Comparing this value with the absolute magnitudes of different reddening and metallicity
solutions at a distance of $1.5-3.0$~Kpc as given in Table~\ref{par} we conclude that if the
Period - Luminosity 
relation of classical Cepheids is valid for \object{V823~Cas}, it is 
either a 1.5~Kpc distant, $T_{eff}=6500$~K object with solar composition
or a cooler, Z=0.01-0.004 metallicity object at about 2 Kpc.
As it was already mentioned in Sect.~4, Cepheids with $P<1$ day are hotter than 6000 K,
thus we conclude that the most probable solution for \object{V823~Cas} 
is E(B-V)=0.37, d=1.5 Kpc, $T_{eff}=6500$~K, $M_V=-0.9$,  $Z=0.02$.
Because of the uncertainties involved, this result has to be taken with
caution.

\begin{table}[]
   \caption{Solutions for dereddening the observed coulours of V823~Cas.}
         \label{par}
\begin{tabular}{c@{\hspace{-1pt}}cccccl}
            \hline\hline
            \noalign{\smallskip}
$E(B-V)$ & $Z$ & $T_{eff}$ &\multicolumn{4}{c}{$M_V$}\\
&&&1.5 & 2.0 & 2.5 &3.0 [Kpc]\\
	    \noalign{\smallskip}
            \hline
	    \noalign{\smallskip}
0.37&0.020&6500&$-0.93$&$-1.56$&$-2.04$&$-2.44$\\
0.19&0.010&5750&$-0.35$&$-0.98$&$-1.46$&$-1.86$\\
0.15&0.004&5500&$-0.22$&$-0.85$&$-1.33$&$-1.73$\\
	    \noalign{\smallskip}
            \hline
\end{tabular}
\end{table}

\section{Summary}

The recent CCD observations of \object{V823~Cas} can be fitted with 
the periods of the 3 lowest radial modes, their harmonics
and their linear combination terms within the
observational accuracy. No indication of nonradial p-modes 
has been found, but the probable existence of a small amplitude signal 
with 1.22~d periodicity needs an explanation.

We present, for the first time, extended multicolour observations
of a multi-mode radial pulsator which is accurate enough to study
the phase and amplitude relations of the radial modes and  the
different order coupling frequency components. These results give a 
deeper insight to the behaviour of the nonlinear processes 
taking place during stellar pulsation.

The observed period ratios of \object{V823 Cas} cannot be consistently 
explained with canonical pulsation and evolution models.
We interpret this anomaly as a consequence of a temporal behaviour
of the periods which may  originate from the very rapid 
evolutionary changes of the stellar parameters and/or a resonance
connection between the radial modes' frequencies ($f_0+f_2=2f_1$),
which might have been fulfilled in the astronomically recent past.  
Both theoretical and observational investigations  are
needed in order to check the reality of this resonance during the 
triple-mode phase of the pulsation, and to examine its effect on 
the observed period values.

If the actual values of the $P_1$ and $P_0$ period change rates
remain unchanged,
a normal $P_1/P_0$ period ratio is going to be reached in a  
short time, in about 10-20\,000 years. This peculiar period change 
behaviour, together with the lapse of the $P_2$ mode, as indicated by its
amplitude decrease,  may also hint 
that in its triple-mode state the periods of \object{V823 Cas} may have 
transient, resonance affected values, which do not reflect the
real physical parameters of the star.

The unique period change behaviour of \object{V823 Cas} warns about the
validity of drawing any conclusion about the evolutionary status and 
the direction of the evolution from the period change rates. 
A similar conclusion was reached from the unexpected period change 
behaviour of the two recently discovered overtone mode triple-mode radial
pulsators in the LMC by \citet{mp}.

\begin{acknowledgements}
We would like to thank Dr S. Antipin for kindly providing us the 
photographic measurements of \object{V823~Cas}. This research has 
made use of the SIMBAD database, operated at CDS Strasbourg, France. 
The financial support of OTKA grants T-043504, T-046207 and T-048961 
is acknowledged.

\end{acknowledgements}

\bibliographystyle{aa}

\end{document}